\documentclass[a4paper,fleqn,usenatbib]{mnras}

\usepackage{newtxtext,newtxmath}

\usepackage[T1]{fontenc}
\usepackage{ae,aecompl}

\usepackage{graphicx}	
\usepackage{amsmath}	

\title[Evolution of Sr and Ba]{Constraints on stellar rotation from the evolution of Sr and Ba in the Galactic halo}

\author[Rizzuti et al.]{ Rizzuti, F.$^{1}$\thanks{E-mail: f.rizzuti@keele.ac.uk}, Cescutti, G.$^{2,3}$, Matteucci, F. $^{2,4,5}$,
  Chieffi, A.$^{6,7,8}$, Hirschi, R.$^{1,9}$, Limongi, M.$^{7,9,10}$, Saro, A.$^{2,3,4,5}$
\\
$^{1}$Astrophysics Group, Lennard-Jones Laboratories, Keele University, Keele ST5 5BG, UK\\
$^{2}$INAF, Osservatorio Astronomico di Trieste, Via Tiepolo 11, I-34143 Trieste, Italy\\
$^{3}$IFPU, Istitute for the Fundamental Physics of the Universe, Via Beirut, 2, I-34151, Grignano, Trieste, Italy\\
$^{4}$Dipartimento di Fisica, Sezione di Astronomia, Universit\`a degli Studi di Trieste, Via Tiepolo 11, I-34143 Trieste, Italy\\
$^{5}$INFN, Trieste, Via Valerio 2, I-34127 Trieste, Italy\\
$^{6}$INAF/IAPS, Via Fosso del Cavaliere 100, I-00133 Roma, Italy\\
$^{7}$INFN. Sezione di Perugia, via A. Pascoli s/n, I-06125 Perugia, Italy\\
$^{8}$Monash Centre for Astrophysics (MoCA), School of Mathematical Sciences, Monash University, Victoria 3800, Australia\\
$^{9}$Kavli IPMU (WPI), The University of Tokyo, Kashiwa, Chiba 277-8583, Japan\\
$^{10}$INAF/Osservatorio Astronomico di Roma, Via di Frascati 33, I-00040 Monte Porzio Catone, Italy}

\date{Accepted XXX. Received YYY; in original form ZZZ}

\pubyear{2020}

\begin{document}
\label{firstpage}
\pagerange{\pageref{firstpage}--\pageref{lastpage}}
\maketitle

\begin{abstract}
Recent studies show that the chemical evolution of Sr and Ba in the Galaxy can be explained if different production sites, hosting r- and s-processes, are taken into account. However, the question of unambiguously identifying these sites is still unsolved. Massive stars are shown to play an important role in the production of s-material if rotation is considered. In this work, we study in detail the contribution of rotating massive stars to the production of Sr and Ba, in order to explain their chemical evolution, but also to constrain the rotational behaviour of massive stars. A stochastic chemical evolution model was employed to reproduce the enrichment of the Galactic halo. We developed new methods for model-data comparison which help to objectively compare the stochastic results to the observations. We employed these methods to estimate the value of free parameters which describe the rotation of massive stars, assumed to be dependent on the stellar metallicity. We constrain the parameters using the observations for Sr and Ba. Employing these parameters for rotating massive stars in our stochastic model, we are able to correctly reproduce the chemical evolution of Sr and Ba, but also Y, Zr and La. The data supports a decrease of both the mean rotational velocities and their dispersion with increasing metallicity. Our results show that a metallicity-dependent rotation is a necessary assumption to explain the s-process in massive stars. Our novel methods of model-data comparison represent a promising tool for future galactic chemical evolution studies.

\end{abstract}




\section{Introduction}
Heavy elements beyond the iron peak are formed through neutron captures \citep{1957RvMP...29..547B}, which are generally divided into two classes: a slow process (s-process) if the timescale for neutron capture is longer than the $\beta$-decay of the freshly synthesized unstable nucleus, and a rapid process (r-process) if it is shorter.
\\For most of the heavy elements we need to take into account both processes in order to explain their production. In the s-process, some peaks of production can be identified (Sr-Y-Zr, Ba-La-Ce-Pr-Nd, and Pb-Bi), linked to the magic neutron numbers 50, 82 and 126, which give particular stability to the nucleus. For this reason, it is interesting to follow the evolution of elements Sr and Ba in the Milky Way, as representative of the first and second peak of the s-process production. 
\\Major sites of s-production are found in low-mass asymptotic giant branch (AGB) stars, with a mass between 1.5–3.0 M$_\odot$ \citep{2009ApJ...696..797C,2011ApJS..197...17C,2010ASSP...16..107K}. A neutron flux is generated through the reaction $^{13}\text{C}(\alpha,\text{n})^{16}\text{O}$, and neutron capture elements can be produced up to Pb-Bi. For a review see \citet{2006NuPhA.777..311S}.
\\But in order to fully explain the s-production, an additional source is needed. Massive stars can produce s-elements through the neutron flux generated by the reaction $^{22}\text{Ne}(\alpha,\text{n})^{25}\text{Mg}$ (“weak s-process”). This mechanism is not very efficient in models without rotation, which can build elements only up to the magic number 50, i.e. Sr-Y-Zr \citep{1992ApJ...387..263R,2003ApJ...592..404L}. The situation changes if rotation is taken into account: the rotation-induced mixing transports chemical species across otherwise unmixed stable radiative zones, thus enabling new nucleosynthesis paths, and it affects the size of the burning core \citep{2008ApJ...687L..95P, 2012A&A...538L...2F, 2013ApJ...764...21C}, so the s-process production and enrichment of neutron capture elements is enhanced. These effects are particularly relevant as the metallicity decreases, because of the large increase of the neutron/seed ratio (i.e. the ratio between the abundance of neutrons and the Fe nuclei), as discussed in \citet{Lim18}. Moreover, there are a number of reasons for which low metallicity massive stars are expected to rotate faster \citep{2002A&A...390..561M,Fris16, Lim18}, therefore in this case the s-process production would be enhanced even more. \citet{Cescutti13}, \citet{Cescutti14} and \citet{Cescutti15} showed that including the s-process from rotating massive stars (RMSs) in chemical evolution models is fundamental in order to explain the heavy element enrichment, in particular of Sr and Ba.
\\On the other hand, for the r-process a large flux of free neutrons is required. The first proposed sites were core collapse SNe or electron capture SNe \citep{1981A&A....97..391T,1991PhR...208..267C}, but they were shown not to have the required entropy and neutron fraction for an efficient r-process activation \citep{2007A&A...467.1227A}. Therefore other sites were proposed, in addition or replacement: neutron star mergers (NSMs; \citealt{1999A&A...341..499R}) or magneto-rotationally driven supernovae (MRD SNe; \citealt{2012ApJ...750L..22W}; \citealt{2015ApJ...810..109N}).
\\Before the NSM event GW170817 observed by LIGO and Virgo \citep{2017PhRvL.119p1101A}, \citet{2014MNRAS.438.2177M} showed that NSMs in a chemical evolution model are able to reproduce the r-process material measured from observations either partially, in a mixed scenario with both SNe II and NSMs, or totally, assuming a very short timescale for the merging after the formation of the binary system (see also \citealt{2004A&A...416..997A,Cescutti15, 2019MNRAS.486.2896S}). More recently, studies include NSMs with time-dependent coalescence timescales and fraction of binary systems \citep{2020arXiv200909534C}. In a similar way, EC SNe and MRD SNe were included in chemical evolution models by \citet{Cescutti13} and \citet{Cescutti14} respectively.
\\Additionally, a direct comparison between NSMs and MRD SNe as source of r-process in a chemical evolution model has been made by \citet{Rizzuti}, where it is shown that the two sites produce very similar results for the evolution of Sr and Ba, if the time delay for the NS merging is fixed to 1 Myr. \citet{Rizzuti} also tested different prescriptions for nucleosynthesis in rotating massive stars \citep{Fris16,Lim18}, showing that the rotational velocity of massive stars should depend on the metallicity, in order to reproduce the observed abundances of Sr and Ba in the Galaxy.
\\In this paper, we intend to analyse in detail the effects of rotation for massive stars on the heavy element nucleosynthesis. Employing a stochastic model for chemical evolution based on \citet{Cescutti08} and \citet{2010A&A...515A.102C}, we not only test the validity of \citet{Fris16} and \citet{Lim18} prescriptions for rotating massive stars to follow the evolution of Sr and Ba, but we also use these results to define new functions which, employed in our model, can effectively describe the rotational velocity of massive stars given their physical parameters.
\\The paper is organized as follows: in Section 2 we describe the observational data we adopted. In Section 3 we present the chemical evolution model. In Section 4 we discuss the nucleosynthesis prescriptions. In Section 5 we introduce new methods to compare the model results to the observational data. In Section 6 the results are presented and in Section 7 some conclusions are drawn.

\section{Observational data}
The model we employ simulates the chemical evolution of the Galactic halo, so we consider data from low metallicity ([Fe/H] from $-4$ to $-1$) Milky Way halo stars taken from various authors (JINA-CEE database, \citealt{2018ApJS..238...36A}). We excluded all upper limits and carbon-enhanced, metal-poor, s-enhanced stars (CEMP-s), since the abundances of s-process elements are affected by mass transfer from an evolved AGB companion \citep{2012MNRAS.422..849B,2012ApJ...747....2L}. For CEMP-s stars we adopt the definition given by \citet{2010A&A...509A..93M}, excluding stars with [C/Fe] $> 0.9$ and [Ba/Fe] $> 1$. The list of the studies considered is displayed in Table~\ref{tab:1}.
\begin{table*}
\centering
\footnotesize
\caption{Sources for observational data abundances.}\label{tab:1}
\begin{minipage}{0.42\linewidth}
\begin{tabular}{lccccc}
\hline
 & Ba & Sr & Y & Zr & La\\
\hline
\citet{Allen12}
& X & X & X& X&X\\
\citet{Aoki02b}
&X  &X  & X& X&\\
\citet{Aoki05}
 & X & X & X&X&X\\
\citet{Aoki07b}
& X & X & &\\
\citet{Aoki13}
& X & X &X &&X\\
\citet{Aoki14}
& X & X & &\\
\citet{Bark05}
& X & X &X &X&X\\
\citet{Bonif09} 
& X & X & &\\
\citet{Cayrel04}
& X & X &X &X&X\\
\citet{2016AA...595A..91C}
& X & X & &\\
\citet{Christlieb04}
& X & X &X &X&\\
\citet{Cohen08}
&X&X&  &  & \\
\citet{Cohen13}
& X & X &X &X&X\\
\citet{Cowan02}
& X & X &X &X&X\\
\citet{Hansen12}
& X & X &X &X\\
\citet{Hansen15}
& X & X & &&\\
\citet{Hayek09}
& X & X & X&X&X\\
\citet{Hollek11}
& X & X & X&X&X\\
\citet{Honda04}
& X & X &X &X&X\\
\citet{Honda11}
& X & X & &\\
\citet{Ivans03}
& X & X &X &&\\
\hline
        \end{tabular}
    \end{minipage}%
    \begin{minipage}{0.42\linewidth}
    \begin{tabular}{lccccc}
\hline
 & Ba & Sr & Y & Zr & La \\
\hline
\citet{Ivans06}
& X & X &X &X&X\\
\citet{Jacobson15}
& X & X & &&\\
\citet{Lai07}
& X & X & &\\
\citet{Lai08}
& X & X & X&X&X\\
\citet{Li15a}
& X & X & X&\\
\citet{Li15b}
& X & X & X&X&X\\
\citet{Mashonkina10}
& X & X & X&X&X\\
\citet{Mashonkina14}
& X & X & X&X&X\\
\citet{Masseron06}
& X & X & X&X&X\\
\citet{McWilliam95}
& X & X & X&X&X\\
\citet{Placco14}
& X & X &X &\\
\citet{Placco15}
& X & X & &&\\
\citet{Preston06}
& X & X &X &X&X\\
\citet{Roederer10}
& X & X & X&X&X\\
\citet{2014AJ....147..136R}
&X&X& X & X & X\\
\citet{2014ApJ...784..158R}
& X & X & X&X&X\\
\citet{Siq14}
& X & X & X&X&X\\
\citet{Spite14}
& X & X &X &X&X\\
\citet{Westin00}
& X & X &X &X&X\\
\citet{Yong13}
& X & X & &\\
\\
\hline
\end{tabular}
    \end{minipage} 
\end{table*}
\\Additionally, the data of the halo star TYC 8442-1036-1 from the work of \citet{2016AA...595A..91C} was taken into account ([Fe/H] $= -3.5$).
\\All the studies normalized the data according to solar abundances taken from \citet{2009ARA&A..47..481A}, enabling a consistent comparison between them.

\section{The chemical evolution model}
The chemical evolution model we adopt for this study is a stochastic model, presented in \citet{2010A&A...515A.102C} and based on the inhomogenous model first developed by \citet{Cescutti08} and on the homogeneous one of \citet{2008A&A...479L...9C}, and later adopted in \citet{Cescutti13} and in other works.
\\The model is intended to reproduce the chemical evolution of the Galactic halo, so it has a time range of 1 Gyr. Inhomogeneities are raised by means of a stochastic process: the halo is considered composed of many cubic regions, which all have the same volume and are independent. The typical volume we chose for the regions is $8\times 10^6$ pc$^3$. Note that this volume is almost 3 times larger than the one taken by \citet{Cescutti08}; in this way, we want to take into account the fact that NSM ejecta can reach larger distances than the other sources of r-process previously used. The total number of volumes which compose the halo was set to 100, in order to produce good statistical results. The dimensions and number of the regions were carefully chosen: the volume is large enough to neglect interactions, but not so large to lose the stochasticity; for larger volumes, the model tends to homogeneous results.
\\For each region, the infall of primordial gas follows the same law as the homogeneous model by \citet{2008A&A...479L...9C}:
\begin{equation}
\dot{G}(t)_\text{inf} = \dfrac{C}{ \sqrt{2\pi}\cdot \sigma_0}\ e^{-(t-t_0)^2/2\sigma_0^2}
\end{equation}
where $t_0$ is 100 Myr, $\sigma_0$ is 50 Myr, and $C$ is $3.2\cdot10^6\ M_{\odot}$. The star formation rate (SFR) $\psi$(t) is defined as
\begin{equation}
\psi(t) =\nu\cdot \dfrac{1}{D^{k-1}} \cdot \rho_\text{gas}(t)^k
\end{equation}
where $\nu$ is the star formation efficiency, here 1.4 Gyr$^{-1}$, $k = 1.5$ the law index, $\rho_\text{gas}(t)$ the amount of the gas inside the volume in $M_{\odot}$, and $D$ is $2\cdot10^6\ M_{\odot}$.
\\Additionally, in this model an outflow is taken into account, considered as gas leaving the system:
\begin{equation}
\dot{G}(t)_\text{out} =W\cdot \psi(t)
\end{equation}
where $W$ is a constant and is set equal to 8 (see \citealt{2008A&A...479L...9C}).
\\For each region, at each timestep, the amount of mass which is transformed into stars $M^\text{new}_\text{stars}$ is fixed to 100 $M_{\odot}$. Then, stars with masses between 0.1 and 100 $M_{\odot}$ are randomly extracted (and weighted according to the initial mass function (IMF) of \citealt{1986FCPh...11....1S}), until the total mass of the newborn stars exceeds $M^\text{new}_\text{stars}$. This cycle is repeated for each region of the halo, so at the end of a timestep all volumes have the same $M^\text{new}_\text{stars}$, but different stellar mass and number distributions. 
After the extractions, the model follows the evolution of the stars, which have different masses therefore different lifetimes (we assume the stellar lifetimes of \citealt{1989A&A...210..155M}), and when they die the ISM is enriched with their ejecta. In this way, the chemical evolution of the Galactic halo is predicted (see \citealt{Cescutti13}).

\section{Nucleosynthesis prescriptions}\label{sec:4}
As mentioned in the Introduction, the presence of many neutron capture elements in the Milky Way is explained by a double production from both r- and s-processes. In particular for barium, works since \citet{1999ApJ...521..691T} and \citet{Cescutti06} indicate a dominant contribution from low-mass AGB stars (and thus s-process) but also a non negligible contribution from r-process. Here we considered an additional source of s-process from rotating massive stars, whose nucleosynthesis is strongly dependent on stellar mass, metallicity and rotational velocity. They have already been included in the studies of \citet{Cescutti13}, \citet{Cescutti14}, \citet{Cescutti15}, \citet{Prantzos18}, \citet{Rizzuti} and \citet{2020MNRAS.491.1832P}, to successfully explain the evolution of different neutron capture elements.
\\Nucleosynthesis by s-process in low mass AGB stars (1.3 - 3 $M_{\odot}$) was taken from the yields of \citet{2009ApJ...696..797C,2011ApJS..197...17C}. Here we used the results from non-rotating stars, but such yields tend to overproduce the neutron capture elements at solar abundance; however, results from rotating stars produce too little neutron capture elements. For this reason, in agreement with \citet{Rizzuti}, we decided to divide the non-rotating yields by a factor of 2, because such a reduction can reproduce the observational data at solar metallicity. We made this choice in order to be consistent with \citet{Rizzuti}, but we do not expect an important effect on our simulation of the Galactic halo, which does not reach high metallicities. Recently, \citet{2020ApJ...897L..25V} suggested that the s-production in rotating AGB stars can be enhanced by including magnetic-buoyancy induced mixing.
\\For the r-process, we employed NSMs as first proposed by \citet{1999A&A...341..499R}. The rate of occurrence and the yields were adopted from the works of \citet{2014MNRAS.438.2177M} and \citet{Cescutti15}, respectively. Their studies prove that r-material can be produced exclusively by NSMs, assuming that neutron stars originate in the mass range of 9 - 50 $M_{\odot}$, the coalescence timescale is fixed and equal to 1 Myr, and the fraction of NS-NS binary systems is 0.018, found from the present-time rate of NS merging by \citet{2004ApJ...614L.137K}. The merging neutron star rate and heavy element production derived by LIGO/Virgo for the event GW170817 have confirmed that these assumptions can explain the r-production in the Milky Way \citep{Matteucci19}. 
\\In some parts of this work we switched the r-process source from NSMs to MRD SNe, with the purpose of making a direct comparison between the two sites, as already done in \citet{Rizzuti}. In employing MRD SNe, we refer to the works of \citet{Cescutti14} and \citet{Rizzuti}, where it was assumed that 10\% of all stars in the mass range 10 - 80 $M_{\odot}$ produce MRD SNe. The adopted r-process yields for Sr and Ba have been obtained from the Solar system r-process contribution, as determined by \citet{2004ApJ...617.1091S}.
\\For the s-process in rotating massive stars, as already introduced by the work of \citet{Rizzuti}, we used alternatively the two different prescriptions of \citet{Fris16} and \citet{Lim18}.
\\\citet{Fris16} produced a large grid of yields using stellar models with dependence on mass, metallicity and rotation. The mass range taken into account is 15 - 40 $M_{\odot}$. Four metallicities are explored: [Fe/H] = $0, -1.8, -3.8,$ and $-5.8$. In our models only the first three metallicities were considered, because for the lowest one (i.e. [Fe/H] $=-5.8$) only a model for 25 $M_{\odot}$ has been computed. Therefore, we decided not to use these results. Instead, we extended the yields from [Fe/H] $=-3.8$ also to lower metallicities.
\\Different initial rotational velocities were taken into account, in relation to the mass and metallicity of the star. For the first two metallicities [Fe/H] = 0 and $-1.8$, we used the results from \citet{Fris16} where the value of standard initial rotation rate over critical velocity was fixed to $v_\text{ini}/v_\text{crit} = 0.4$. Keeping this ratio constant, the resulting average equatorial rotation velocity on the main sequence $\langle v \rangle_\text{MS}$ increases with decreasing metallicity. E.g. for 15 - 20 $M_{\odot}$ stars at solar metallicity, $\langle v \rangle_\text{MS}$ corresponds to 200 - 220 km/s.
\\For the metallicity [Fe/H] $= -3.8$, in order to account for a stronger s-production, we decided to use results which provide a faster rotation, i.e. a higher ratio $v_\text{ini}/v_\text{crit} = 0.5$, and a lower $^{17}\text{O}(\alpha,\gamma)$ rate (one tenth of the standard choice, i.e. \citealt{1988ADNDT..40..283C}). The only model produced by \citet{Fris16} with these assumptions takes into account only the stellar mass of 25 $M_{\odot}$, but we decided to extend these results to other masses. We computed for each element a ratio between the yields of 25 $M_{\odot}$ obtained from the fast rotator model and the ones from the standard model, and then applied the resulting scale factors to the other models with metallicity [Fe/H] $= -3.8$ and masses 15, 20, and 40 $M_{\odot}$ (as also done in \citealt{Cescutti13}, \citealt{Rizzuti}).
\\We display in Table~\ref{tab:2} the \citet{Fris16} models used in our work with their features.
\begin{table}
\centering
\footnotesize
\caption{Model parameters adopted for our work from \citet{Fris16}: initial mass, model label, initial ratio of surface velocity to critical velocity, time-averaged surface velocity during the MS phase, metallicity.}\label{tab:2}
\begin{tabular}{lcccr}
\hline
\hline
Mass (M{$_\odot$}) & Model & $v_{\text{ini}}/v_{\text{crit}}$ & $\langle v \rangle_{\text{MS}}$ (km/s) & [Fe/H] \\
\hline
15 & A15s4 & 0.4 & 200 & 0.0 \\
   & B15s4 & 0.4 & 234 & $-1.8$ \\
   & C15s4 & 0.4 & 277 & $-3.8$ \\
20 & A20s4 & 0.4 & 216 & 0.0 \\
   & B20s4 & 0.4 & 260 & $-1.8$ \\
   & C20s4 & 0.4 & 305 & $-3.8$ \\
25 & A25s4 & 0.4 & 214 & 0.0 \\
   & B25s4 & 0.4 & 285 & $-1.8$ \\
   & C25s4 & 0.4 & 333 & $-3.8$ \\
   & C25s5b$^a$ & 0.5 & 428 & $-3.8$ \\
40 & A40s4 & 0.4 & 186 & 0.0 \\
   & B40s4 & 0.4 & 334 & $-1.8$ \\
   & C40s4 & 0.4 & 409 & $-3.8$ \\
\hline
\multicolumn{5}{l}{$^a$ Models calculated with a lower $^{17}$O($\alpha,\gamma$).}
\end{tabular}
\end{table}
\\On the other hand, the work of \cite{Lim18} produced a grid of yields based on a mass range of 13 - 120 $M_{\odot}$, and four metallicities: [Fe/H] $= 0, -1,-2,$ and $-3$. This grid was computed for three different stellar rotational velocities, namely 0 km/s (non-rotating), 150 km/s and 300 km/s. In this way, for each star with a certain mass and metallicity it is possible to choose one of the three rotational speeds, as needed.
\\It is important to note the differences between the two works. The models of \citet{Lim18} have been computed up to the pre-SN stage, and their explosive nucleosynthesis has been taken into account by means of induced explosions, while models in \citet{Fris16} stop at the beginning of the O-core burning. In the models we employed from \citet{Lim18}, the amount of matter effectively ejected is the one lost by the star through stellar wind during the pre-SN evolution, plus the one ejected during the explosion. The mass cut between the collapsing core and the ejected envelope has been fixed in such a way that the ejecta contains 0.07 $M_{\odot}$ of $^{56}$Ni, a typical value observed in the spectra of core collapse SNe. In fact, among the \citet{Lim18} sets developed for this scenario, we used here Set F, which is the one where each mass is considered to eject 0.07 $M_{\odot}$ of $^{56}$Ni.
\\Concerning one of the most relevant aspects of the two works, the assumption of rotation, on the one hand \citet{Fris16} produced models where stars have no rotation or rotate with a specific velocity which depends on their mass and metallicity, while on the other hand \citet{Lim18} adopted the same three velocities 0, 150 and 300 km/s for all stars, producing results where it can be possible to choose the stellar velocity. In this way, data from \citet{Fris16} can be used directly in an evolution model which takes into account stellar rotation, as done by \citeauthor{Cescutti13} (\citeyear{Cescutti13}, \citeyear{Cescutti14}, \citeyear{Cescutti15}) using \citet{2012A&A...538L...2F} and by \citet{Rizzuti} using \citet{Fris16}. On the contrary, a model can employ the yields of \citet{Lim18} only making some assumptions about the distribution of stellar velocity, as in \citet{Prantzos18} and \citet{Rizzuti}.
\\In particular, the work of \citet{Rizzuti} assumed that all stars, regardless of their mass or metallicity, rotate with the same speed, but none of the tested velocities was able to explain the data over the entire range of metallicity. In this study, we relax this approximation and allocate to massive stars a new distribution of rotational speed. The main focus of this study is to constrain this distribution.
\\Finally, the iron yields from core collapse SNe were adopted from \citet{2006ApJ...653.1145K}, which are the same as used by \citet{2014MNRAS.438.2177M} and \citet{Rizzuti}. It could have been possible for us to use the ones from \citet{Lim18} instead, but the two works lead to very similar results, and we choose to be consistent with \citet{Rizzuti} which already used them in their homogeneous model.

\section{Methods for model-data comparison}\label{sec:5}
There are many advantages to employ the stochastic chemical evolution model, which can explain the possible spread observed in heavy element abundances, but some difficulties arise when comparing the model results to the observations. Previous studies making use of a stochastic model (\citealt{Cescutti08}, \citeauthor{Cescutti13} \citeyear{Cescutti13}, \citeyear{Cescutti14}, \citeyear{Cescutti15}) visually compared model results and observations to draw qualitative conclusions. Their main purpose was testing prescriptions which deeply affect the shape of the resulting plot, so it was possible to visually check if the assumptions were in agreement with the data or not.
\\In this work, we are comparing prescriptions from different authors regarding the same phenomena, so we are not expecting the results to differ much from one another. We are also interested in slightly adjusting some parameters in our model, resulting in small differences between the plots. In this way, it may not be possible to see immediately from the graphs which assumption or parameter in the model is the best at reproducing the data.
\\For this reason, we propose here a new method of comparison between stochastic results and observational data, adopting an algorithm which can produce a unique numerical value estimating the efficiency of the model in reproducing the data. We note that up to date there are no studies in the literature which apply a comparison method to stochastic chemical evolution models.
\\We choose to employ the likelihood function, which estimates the goodness of fitting a given distribution to a sample of data. The testing is conducted in this way. We consider here the model output as the fitting function and the observational data as one of its possible realizations. Therefore, we use the fact that the likelihood function in a given point is equal to the value of the distribution in that point, so we define the index $L$ as the sum of the model values over all the data points:
\begin{equation}\label{eq 4}
L=-\sum_{data} \log \left(model\left[data\right]\right)
\end{equation}
with a logarithm inside the summation, which smooths the gradient of the index without altering its monotonicity. Also, a minus sign is added to the formula: the likelihood represents the probability that the data were drawn from the model, which is something we want to maximize; reversing the sign makes $L$ an index to minimize.
\\When computing the $L$ index, we faced the problem of treating the data which are not covered by the model, i.e. points where the model is equal to zero. In fact, the formula above is not applicable in this case, since the logarithm produces an infinity. To solve this problem we decided to manually assign a finite number when an infinity is reached. Noticing that the typical likelihood value in our model is about 0 - 10, we replaced the infinity with 100 to penalize points which are not covered by the model. We observed that using larger values as a replacement does not change the behaviour of $L$.
\\Another problem is that the result of the stochastic model is a sample of points with associated weights and not a continuous function, so assigning a value to each point is not straightforward. We decided to use here a normalized 2D-histogram of the stochastic model. In this way, $L$ is computed from the values of the histogram bins where the data points fall. We chose a binning of 0.2 dex in the metallicity space, which is the usual error bar in observational measurements.
\\We compute the $L$ index to compare results coming from different assumptions in the stochastic model: according to the likelihood-ratio test, the model with the lowest $L$ (as defined above) is the one which can best reproduce the observations. But we can also use this method to estimate free parameters in the stochastic model. In order to do so, we apply the maximum likelihood method: we perform a random sampling in the parameter space, run the model with the chosen parameters and compute the $L$ index. The best choice for the free parameters is given by the minimum of the resulting curve, and from its shape we can identify the associated error and the correlation between more parameters.
\\In particular, if we assume that the data were generated by a Gaussian process, we can express the likelihood function $L(\theta)$ of the free parameter $\theta$ as a Gaussian:
\begin{equation}
L(\theta)=L_\text{max}\ e^{-\frac{(\theta-\hat{\theta})^2}{2 \sigma_{\theta}^2}}
\end{equation}
where $\hat{\theta}$ is the estimate for $\theta$, and $\sigma_{\theta}$ its standard deviation. From this we have:
\begin{equation}
\text{ln}L(\theta)=\text{ln}L_\text{max}-\frac{(\theta-\hat{\theta})^2}{2 \sigma_{\theta}^2}
\end{equation}
therefore, in order to find the error on the estimated parameter with a confidence level of 68\%, we impose $\theta=\hat{\theta}\pm \sigma_{\theta}$ and look for the values of $\theta$ which satisfy $\text{ln}L(\theta)=\text{ln}L_\text{max}-1/2$. We recall that we defined the $L$ index in (\ref{eq 4}) with the logarithm and changing the sign, so in our case we want to solve $L=L_\text{min}+1/2$. 
\\This method is also valid for more than one parameter (multi-dimensional likelihood function), but in this case the equations above return a confidence ellipse (or a multi-dimensional ellipsoid), which gives us information about the correlation between parameters.
\\These comparison methods are successfully applied in the following Sections~\ref{sec:6.1} and~\ref{sec:6.2}, and they prove to be extremely useful in comparing and choosing different assumptions in our stochastic model.

\section{Results}
\subsection{Comparison methods for r-process site choice}\label{sec:6.1}
\begin{figure*}
\centering
\footnotesize
\includegraphics[trim={0cm 0 0cm 0cm},clip,width=1.\textwidth]{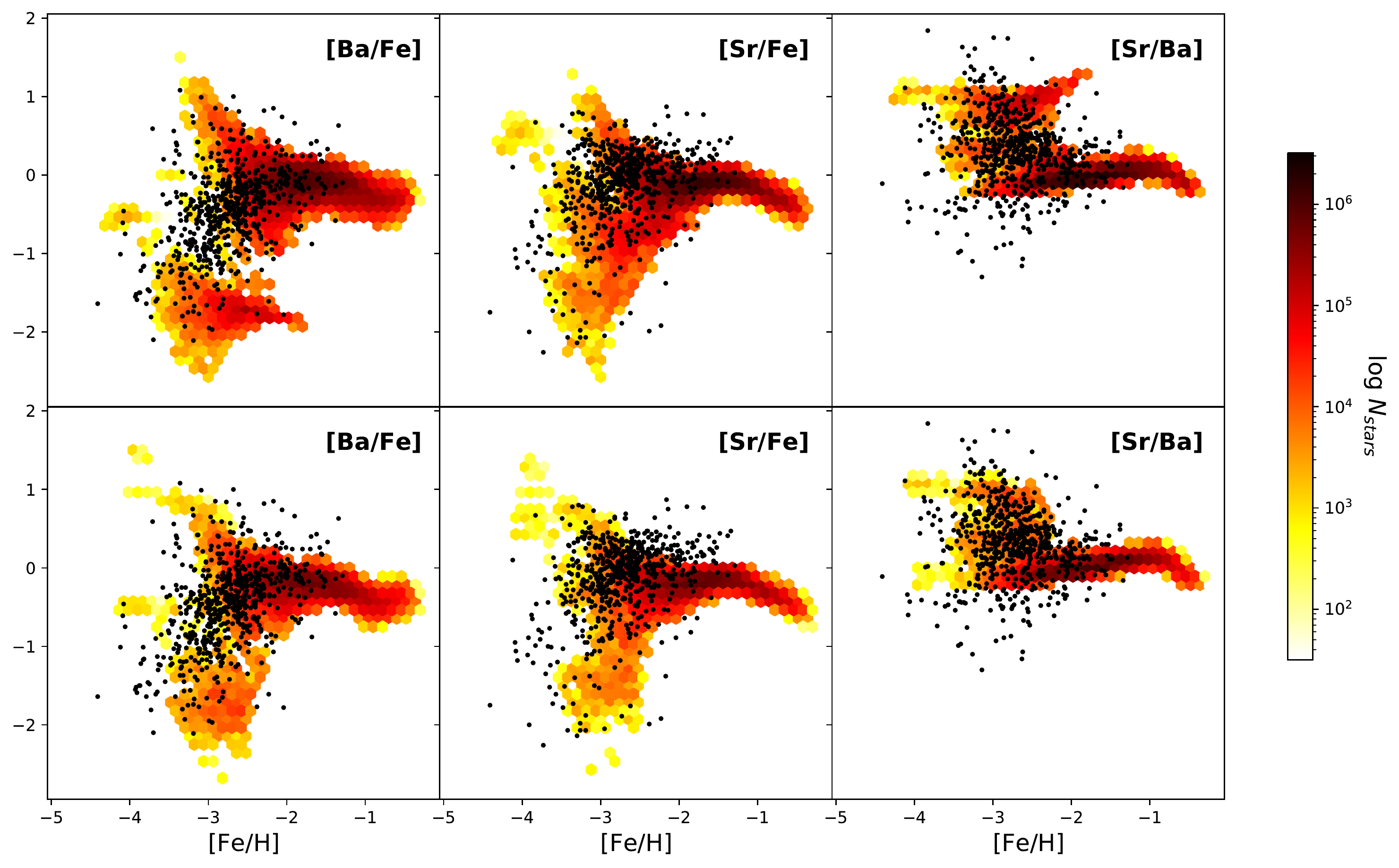}\put(-484,152){\textit{\Large MRD SNe}}\put(-484,295){\textit{\Large NSMs}}
\caption{From left to right, the three ratios  [Ba/Fe], [Sr/Fe] and [Sr/Ba] versus [Fe/H]. The shaded area displays the number of simulated long-living stars for the model on a logarithmic scale (see colorbar). Superimposed on the shaded area, we show the abundance ratios for halo stars (black dots). The first row (\textit{NSMs}) is the model using NSMs as source of r-process and yields from \citet{Fris16} for RMSs. The second row (\textit{MRD SNe}) is the model using MRD SNe as source of r-process and yields from \citet{Fris16} for RMSs.}\label{fig:1}
\end{figure*}

We first apply the methods for model-data comparison developed in Section~\ref{sec:5} in order to directly compare two different prescriptions for r-process, MRD SNe and NSMs, and check which one can best reproduce the observations when employed in the stochastic model. In this case, we use the prescriptions for s-process in rotating massive stars from \citet{Fris16}. For the r-process, we use the prescriptions for magneto-rotationally driven supernovae and neutron star mergers as described in Section~\ref{sec:4}. Note that a version of the stochastic model with MRD SNe and \citet{2012A&A...538L...2F} yields for RMSs has already been presented in \citet{Cescutti14}.
\\We present here the abundance ratios [Ba/Fe], [Sr/Fe] but also [Sr/Ba], which provides differential information about the production of the two elements. In Fig.~\ref{fig:1} we show the results of the two versions of the stochastic model using as source of r-process NSMs in the first row, and MRD SNe in the second row.
\\We can see that employing NSMs or MRD SNe as r-process source produces very similar results, so both scenarios can be considered valid. More generally, the stochastic model using yields from \citet{Fris16} for RMSs covers the data well at medium-high metallicities, but it has some problems at lower ones. For ratios [Ba/Fe] and [Sr/Fe] the model predicts many stars at low metallicity below [X/Fe] $< -1$, but the observations do not support this result. It is worth to underline that it is problematic to measure Sr and Ba if their abundances are very low, so there could be an observational bias. 
On the other hand, the ratio [Sr/Ba] seems to cover most of the data at all metallicities, even though there are some observations not explained by the model. We recall that \citet{2011Natur.472..454C} introduced the idea of explaining the observed spread in [Sr/Ba] with production from rotating massive stars.
\\We can now apply the methods of Section~\ref{sec:5} to the two models, in order to check which one is closer to the data. We chose to estimate the goodness of the model fitting the data in the metallicity space, which means using the plots in Fig.~\ref{fig:1}. In particular, we compute $L$ as defined in equation (\ref{eq 4}) for each of the three graphs [Ba/Fe], [Sr/Fe], [Sr/Ba], and then we sum them to obtain a unique $L$ for each model. We decided to use also the [Sr/Ba] graph to avoid losing any information, since we cannot exclude that the three ratios are not fully independent. In case they are, this only repeats the same information in all computations, so the relative comparison is not altered. 
\\As an example, for the first case (\textit{NSMs}) we use the three plots in the first row of Fig.~\ref{fig:1}, considering all the data (black dots) and using the normalized value of the model (density plot) in these points into equation (\ref{eq 4}) to obtain $L([Ba/Fe])=11257$, $L([Sr/Fe])=8168$ and $L([Sr/Ba])=13990$. We then sum them to obtain the final $L=33415$ for the model with NSMs.
\\We compute the $L$ index as described above, and the results are:
\begin{center}
\begin{tabular}{lc}
\multicolumn{1}{c}{model} & $L$\\
\hline
\citet{Fris16} + NSMs&33415\\
\citet{Fris16} + MRD SNe & 36601\\
\end{tabular}
\end{center}
As we can see, even if it is not immediately visible from the plots, the method returns a smaller $L$ for employing NSMs as source of r-process. According to the likelihood-ratio test, this means that using NSMs in our stochastic model produces an output which is closer to the observational data. To check if this result is consistent and significant, we run again this test at the end of Section~\ref{sec:6.2} with a different prescription for RMSs. In any case, assuming NSMs as source of r-process allows to correctly reproduce the observations, so we decide to keep this fixed in our model and focus on the s-process in the following sections.

\subsection{A velocity distribution for rotating massive stars}\label{sec:6.2}
\begin{figure*}
\centering
\footnotesize
\includegraphics[trim={0cm 0cm 0cm 0cm},clip,width=1.\textwidth]{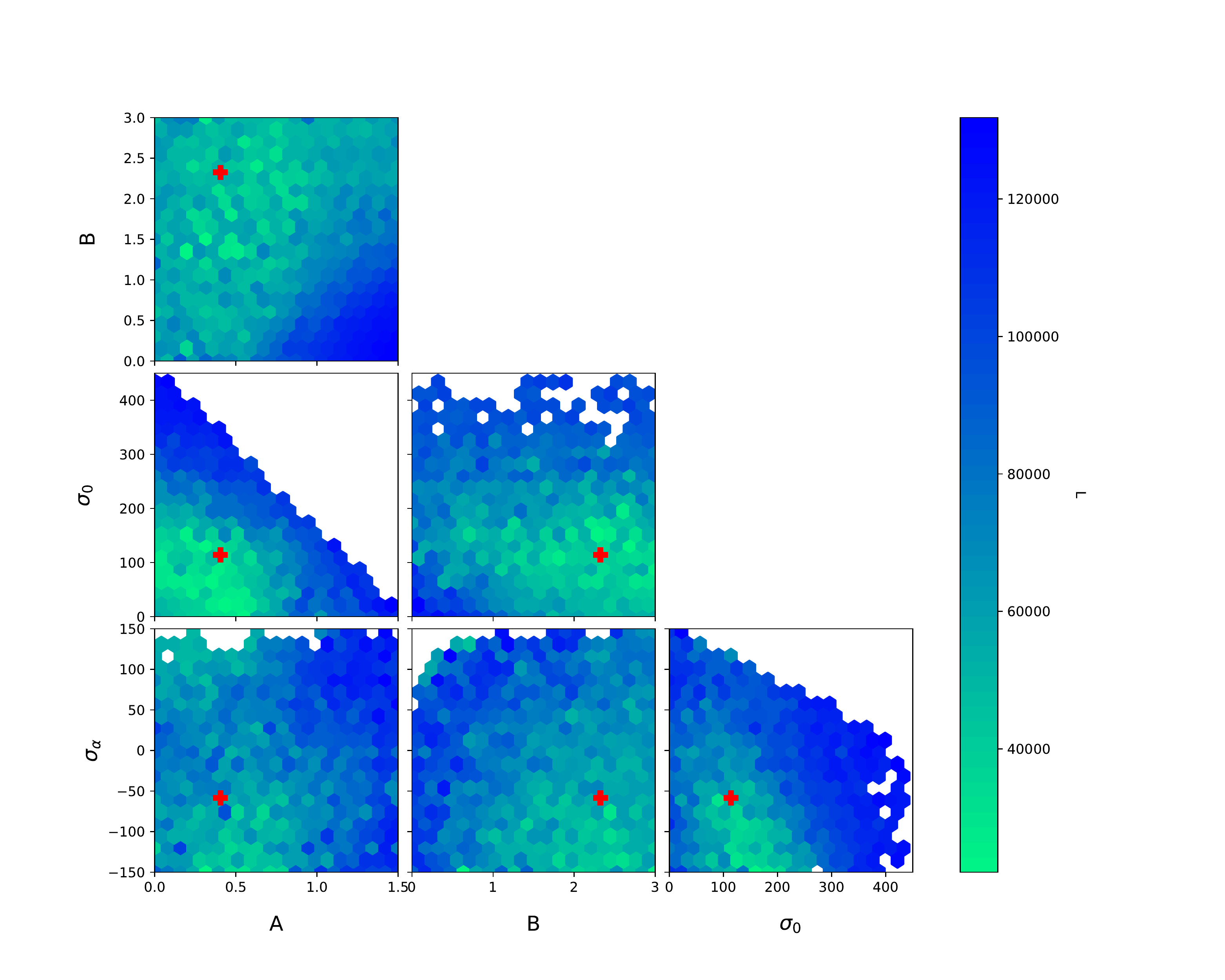}
\caption{Projections of the multi-dimensional surface for the $L$ index as defined in Section~\ref{sec:5}, built from chemical evolution simulations in the space of parameters $A, B, \sigma_0,\sigma_{\alpha}$. Colour scale for $L$ is associated. Red crosses are the coordinates for $L$ minimum.}\label{fig:2}
\end{figure*}

\begin{figure*}
\centering
\footnotesize
\includegraphics[trim={0cm 0 0cm 0cm},clip,width=1.\textwidth]{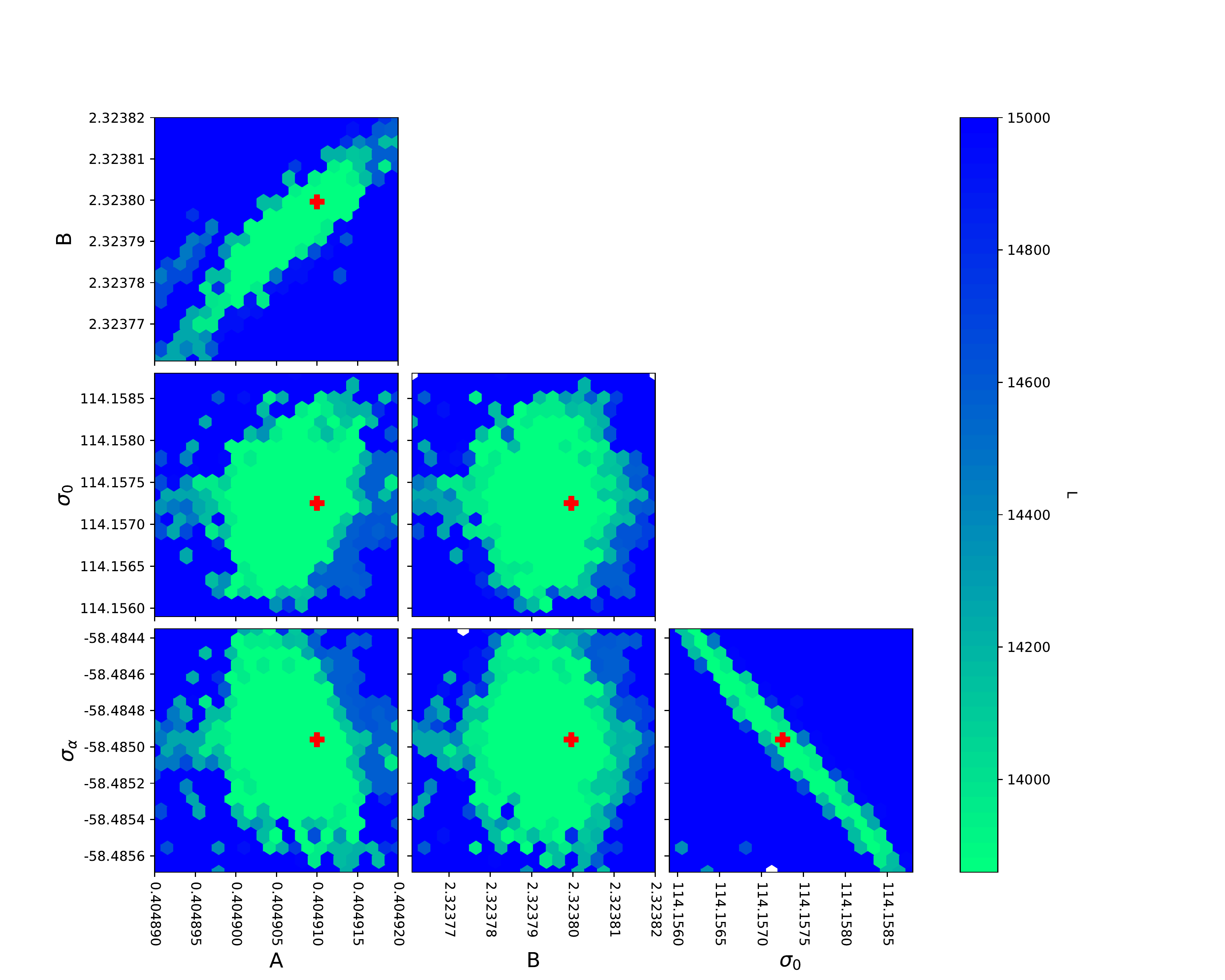}
\caption{Same as Fig.~\ref{fig:2}, but in a smaller range of parameters $A, B, \sigma_0,\sigma_{\alpha}$, focusing around the $L$ minimum (red crosses).}\label{fig:3}
\end{figure*}

\begin{figure*}
\centering
\footnotesize
\includegraphics[trim={0cm 0 0cm 0cm},clip,width=0.8\textwidth]{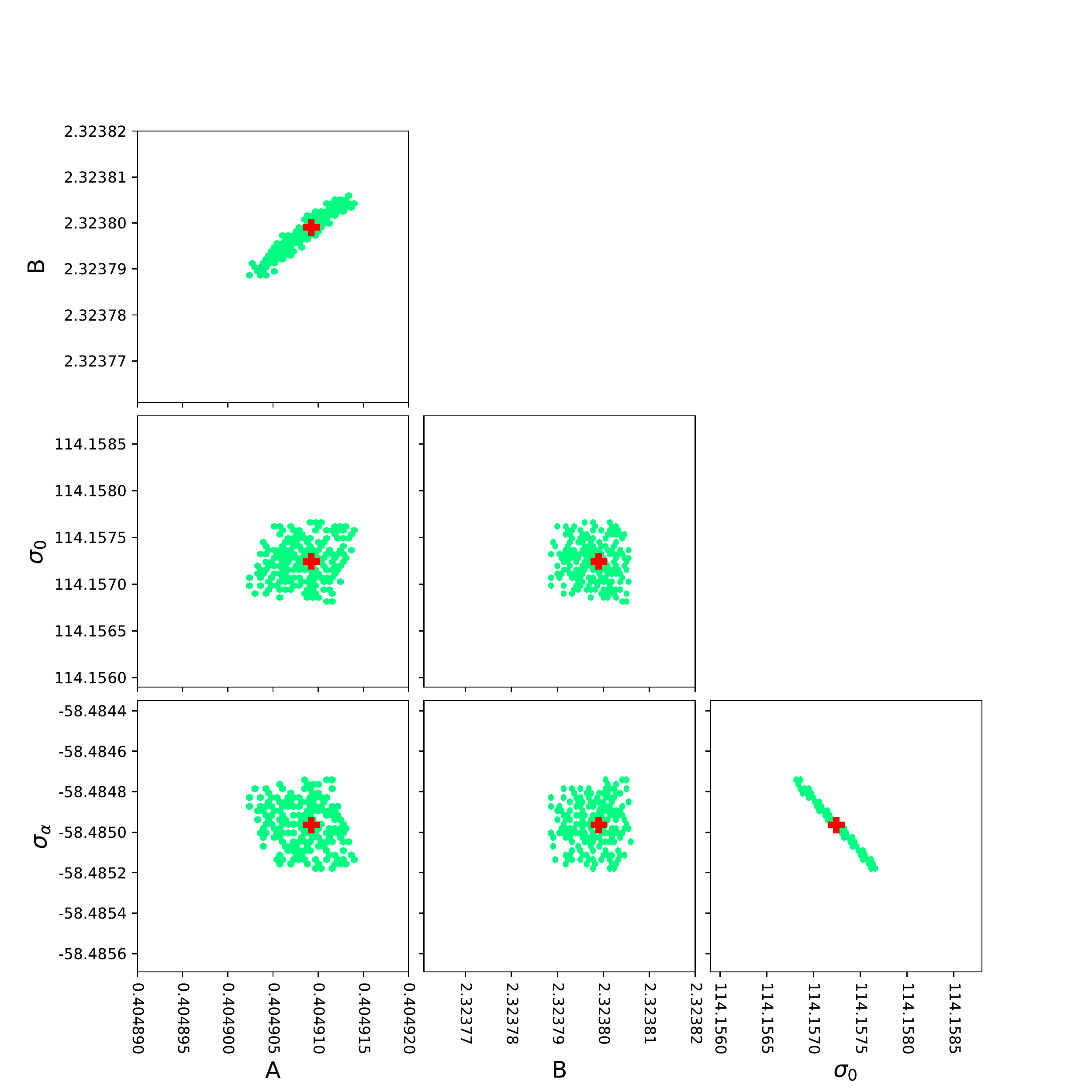}
\caption{Same as Fig.~\ref{fig:3}, but taking only the points where $L=L_\text{min}+1/2$, which define the confidence ellipsoid for the parameter estimation, as described in Section~\ref{sec:5}.}\label{fig:4}
\end{figure*}

\begin{figure*}
\centering
\footnotesize
\includegraphics[trim={0cm 0 0cm 0cm},clip,width=0.67\textwidth]{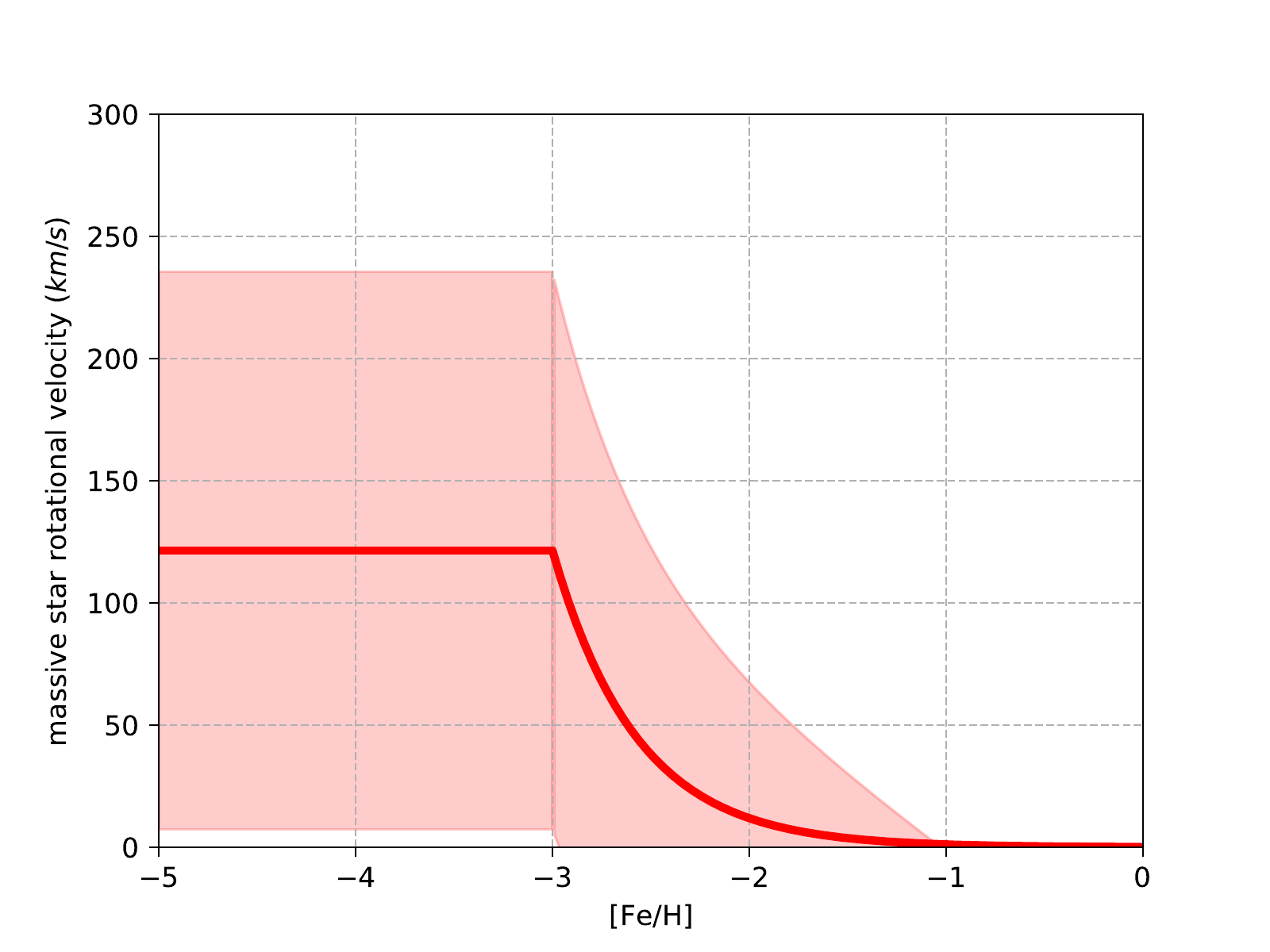}
\caption{Rotational velocity for massive stars versus metallicity. The red line is the mean velocity $\mu$ described by function (\ref{eq:1}), the red shaded area is the Gaussian width $\sigma$ described by function (\ref{eq:2}).}\label{fig:5}
\end{figure*}

\begin{figure*}
\centering
\footnotesize
\includegraphics[trim={0cm 0 0cm 0cm},clip,width=1.\textwidth]{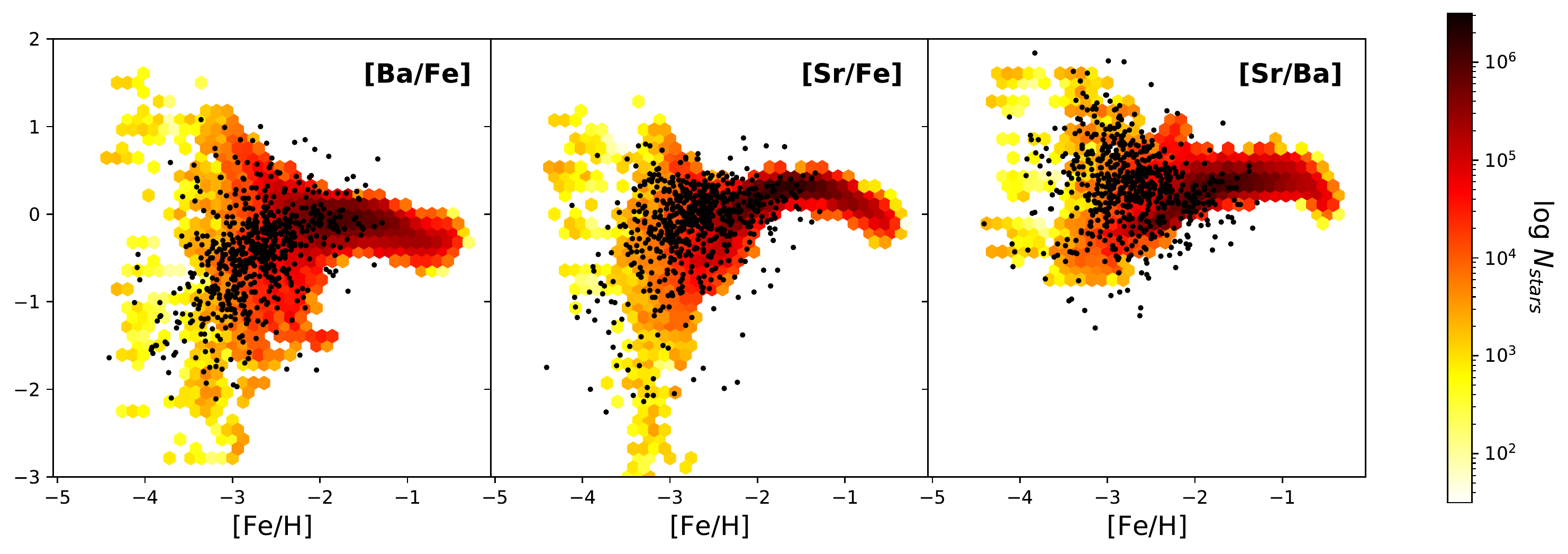}
\caption{Same as Fig.~\ref{fig:1}, but for the stochastic model using NSMs as source of r-process, and yields from \citet{Lim18} for massive stars with rotational velocities following a Gaussian curve with $\mu$ and $\sigma$ described by functions (\ref{eq:1}) and (\ref{eq:2}) respectively.}\label{fig:6}
\end{figure*}

As we introduced in Section~\ref{sec:5}, it is possible to use the methods developed for model-data comparison to estimate free parameters in the model. We proceed now to use the Sr and Ba observations as a constraint for our model in order to study how the rotational velocity of massive stars should depend on the stellar metallicity.
\\As showed by \citet{Rizzuti}, a constant rotational velocity for RMSs cannot explain the s-production of heavy elements Sr and Ba. For this reason, we expect massive stars to rotate faster at lower metallicities, in agreement with the studies of \citet{Fris16}, \citet{Prantzos18} and \citet{Rizzuti}.
\\We assume that rotational velocities of massive stars follow a Gaussian  probability distribution. This assumption should reproduce the real case scenario, where stars have different velocities randomly scattered around a central value. We chose to describe the centre of the Gaussian curve with an exponentially decreasing function of the stellar metallicity:
\begin{equation} \label{eq 7}
\mu = \left\{
 \begin{array}{ll}
 300\cdot A\cdot \exp\left\{-\ B\cdot (\text{[Fe/H]} + 3)\right\}\ km/s & \quad \text{for  [Fe/H]}  \geq  -3 \\
 300\cdot A \ \ km/s & \quad \text{for  [Fe/H]} < -3
 \end{array}
\right.
\end{equation}
where $A$ and $B$ are the free parameters which describe the curve. In this function a step is present: for [Fe/H] $< -3$ the velocity is constant, fixed to the value it has at [Fe/H] $= -3$. We made this choice because the lowest metallicity computed in \citet{Lim18} is [Fe/H] $= -3$, and our model extends these data also to lower metallicities. We chose an exponential function to describe the rotation of massive stars, as already done by \citet{Prantzos18}, since it is monotonic, it does not reach negative values, and it approaches asymptotically a constant value. 
\\Also for the width of the Gaussian distribution we assume a dependence on the stellar metallicity. We describe the Gaussian $\sigma$ with a generic linar function of the metallicity:
\begin{equation} \label{eq 8}
\sigma= \left\{
 \begin{array}{ll}
\sigma_0+\sigma_{\alpha} \cdot \left(\text{[Fe/H]}+3\right) & \quad \text{for  [Fe/H]}  \geq  -3 \\
\sigma_0 & \quad \text{for  [Fe/H]} < -3\\
0 & \quad \text{for } \sigma_0+\sigma_{\alpha} \cdot \left(\text{[Fe/H]}+3\right)<0
 \end{array}
\right.
\end{equation}
where the free parameters which characterize the function are $\sigma_0$ and $\sigma_{\alpha}$. In addition to the step in metallicity defined also for $\mu$, we assume $\sigma$ to be zero in case it reaches negative values.
\\For this study, we chose to employ the yields of \citet{Lim18} for the s-process in RMSs, from which we recall it is possible to choose the velocity between 0, 150 or 300 $km/s$. Since the functions we defined for $\mu$ and $\sigma$ are continuous, when an intermediate velocity is extracted we compute the new yields by means of a linear interpolation between the existing grids. Here, we use NSMs as source of r-process, as described in Section~\ref{sec:4}.
\\In order to fully characterize the defined functions, we need to estimate the free parameters $A, B, \sigma_0,\sigma_{\alpha}$. To do so, we make use of the maximum likelihood method as described in Section~\ref{sec:5}. We proceed in the following way. The parameter space is investigated with a random sampling. For each sample of four parameters ($A, B, \sigma_0,\sigma_{\alpha}$), the stochastic model is run with RMSs following the velocity distribution described by equations (\ref{eq 7}) and (\ref{eq 8}). Then, from the model results and the observations we compute the $L$ index corresponding to that model, as defined in Section~\ref{sec:5}. As we did in the previous section, we compute $L$ from the plots in the metallicity space and using each of the three graphs [Ba/Fe], [Sr/Fe], [Sr/Ba], summing the three indices to obtain a unique $L$ for each model.
\\In order to focus only on the most interesting scenarios, we impose some boundary conditions to the sampling of parameters. The conditions were chosen in a way that rotational velocities stay positive and never larger than 450 $km/s$, which is about the fastest rotation considered by \citet{Fris16} (see Table~\ref{tab:2}).
\\Following this process, from the random sampling a surface in the multi-dimensional parameter space is created for $L$: the more accurate the sampling will be, the more detailed the surface will appear. Our sampling is composed of ${\sim}3200$ extractions, which allow us to have a well-defined surface. We show in Fig.~\ref{fig:2} the $L$-surface projected onto the 2D planes given by all combinations of the four parameters.
A colour scale is associated with the plot, representing the value of $L$ computed for each simulation. We recall that, according to the maximum likelihood estimation described in Section~\ref{sec:5}, lower values for the $L$ index (toward the green, in our plot) correspond to better estimates of the free parameters.
\\In the plot of Fig.~\ref{fig:2}, the white areas without points are the effect of imposing boundary conditions to the parameters, and we notice that near the boundaries the $L$ index assumes higher values (toward the blue), which means that the boundary conditions were chosen reasonably.
\\Finally, the estimate of the four parameters $A, B, \sigma_0,\sigma_{\alpha}$ is given by the coordinates of the minimum $L$ (i.e. $L=13859.79$) plotted as red crosses in Fig.~\ref{fig:2}:
\begin{center}
\begin{tabular}{lcr}
$A$ &=& 0.40490923\\
$B$ &=& 2.32379901\\
$\sigma_0$ &=& 114.157244\\
$\sigma_{\alpha}$ &=& $-58.484965$\\
\end{tabular}
\end{center}
In order to determine the errors and correlations for these parameters, we study in detail the shape of the valley surrounding the $L$ minimum. We run more simulations in a restricted area around the coordinates of the minimum and we obtain the plot in Fig.~\ref{fig:3}.
We can clearly see the features which characterize the minimum valley, with some strong indications of correlation between couples $A,B$ and $\sigma_0,\sigma_{\alpha}$. 
\\In order to obtain numerical values for the error and correlation, we define the confidence ellipsoid taking all points in Fig.~\ref{fig:3} with $L=L_\text{min}+1/2$, as described in Section~\ref{sec:5}. In this way, we plot in Fig.~\ref{fig:4} the projections of the resulting ellipsoid onto the grid of parameters. 
From this plot it is easy to measure the error for each parameter, taken as the extension of the ellipses, while from their shape we can deduce the correlation between parameters. We compute the correlation coefficient $\rho$ according to the formula:
\begin{equation}
\rho_{x,y}=\dfrac{\sum_i(x_i-\hat{x})(y_i-\hat{y})}{\sqrt{\sum_i(x_i-\hat{x})^2}\sqrt{\sum_i(y_i-\hat{y})^2}}
\end{equation}
where $\hat{x}$ and $\hat{y}$ are the means of two parameters $x,y$.
\\In this way, we obtain the errors on the estimates:
\begin{center}
\begin{tabular}{lccc}
&\text{estimate}&\text{lower error}&\text{upper error}\\
\hline
$A$ & 0.40490923&$-$6.83e$-$06&$+$4.76e$-$06\\
$B$ & 2.32379901&$-$10.39e$-$06&+6.95e$-$06\\
$\sigma_0$ & 114.157244&$-$4.29e$-$04&$+$4.16e$-$04\\
$\sigma_{\alpha}$ & $-58.484965$&$-$2.14e$-$04&$+$2.23e$-$04\\
\end{tabular}
\end{center}
and the correlation between different parameters:
\begin{center}
\begin{tabular}{cccc}
$\rho_{x,y}$&$A$&$B$&$\sigma_0$\\
\hline
$B$ & 0.958&&\\
$\sigma_0$ & 0.191&$-0.078$&\\
$\sigma_{\alpha}$ &$-0.220$ &0.044&$-0.998$\\
\end{tabular}
\end{center}
Concerning the correlation between parameters, we can see that $\rho(A,\sigma_0), \rho(A,\sigma_{\alpha}),\rho(B,\sigma_0)$ and $\rho(B,\sigma_{\alpha})$ have such small values that their relationship is not significant. On the other hand, $\rho(A,B)$ and $\rho(\sigma_0,\sigma_{\alpha})$ are close to 1 and $-1$, respectively, meaning that these two couples are fully correlated and anti-correlated, respectively.
\\It is noticeable that the errors on the parameter estimates are particularly small. We recall that with the methods we developed in Section~\ref{sec:5} and applied here, we constrain only the free parameters which describe the rotation of massive stars in our model, while keeping fixed every other parameter and prescription. In this way, we do not take into account the uncertainties which characterize the other aspects of our model (e.g. nucleosynthesis prescriptions, interpolation over mass and metallicity) that should affect also the errors on the parameter estimates. For this reason, we realistically expect a larger uncertainty on the results of the estimation than the one obtained above.
\\In order to have an idea on how uncertainties could affect the errors on the parameter estimates, we decided to compute again the $L$ index for our best model, with the parameters estimated above, but giving to the observational data a simulated Gaussian error. We added to each data an error randomly extracted from a Gaussian curve with $\sigma=0.1$ dex, so in this way $2\sigma$ has the same value as the binning in the histograms we build for the model results. Running this test more times for stochasticity, we find that $L$ varies from the minimum we found ($L=13860$) up to 15824, on average. Taking this $L$ value to define the errors on the estimated parameters, we find that more realistic errors are:
\begin{center}
\begin{tabular}{lccc}
&\text{estimate}&\text{lower error}&\text{upper error}\\
\hline
$A$ & 0.405&$-$0.353&$+$0.140\\
$B$ & 2.324&$-$1.312&+0.389\\
$\sigma_0$ & 114.2&$-$34.6&$+$69.4\\
$\sigma_{\alpha}$ & $-$58.5&$-$89.0&$+$1.5\\
\end{tabular}
\end{center}
This shows that uncertainties can have an effect on evaluating the errors for parameter estimates. Even if the errors are larger, this does not change the behaviour of the parameters we are trying to constrain ($\mu,\sigma$).
\\We can conclude that rotational velocity in massive stars is well described by a Gaussian curve whose centre and width depend on the stellar metallicity according to the functions: 
\begin{equation} \label{eq:1}
\mu = \left\{
 \begin{array}{ll}
 300\cdot 0.405\cdot \exp\left\{-\ 2.324\cdot (\text{[Fe/H]} + 3)\right\}\ km/s \\
\hfill \text{for  [Fe/H]}  \geq  -3 \\
 300\cdot 0.405\ \ km/s \\
\hfill \text{for  [Fe/H]} < -3
 \end{array}
\right.
\end{equation}
\begin{equation} \label{eq:2}
\sigma= \left\{
 \begin{array}{ll}
114.2-58.5 \cdot \left(\text{[Fe/H]}+3\right) & \hfill  \text{for $-3 \leq$ [Fe/H]}  \leq  -1 \\
114.2 & \hfill \text{for  [Fe/H]} < -3\\
0 & \hfill \text{for  [Fe/H]} \geq -1
 \end{array}
\right.
\end{equation}
We can see these functions represented in Fig.~\ref{fig:5}, where we show how rotational velocities of massive stars behave with the metallicity. The red line is the expected centre of the distribution, described by function (\ref{eq:1}), while the red shaded zone is the $1\sigma$ Gaussian dispersion described by function (\ref{eq:2}). As we can see, massive stars at low metallicity can reach fast rotation thanks to the high dispersion around the mean value, but when the metallicity increases not only the mean rotational velocities decrease, but also their dispersion. In particular, for [Fe/H] $>-1$ functions predict little rotation and no dispersion. We recall that the stochastic model we employed is intended to reproduce only the Galactic halo, and there are few observations at high metallicity which can be used to constrain the parameters. Therefore, the conclusion that all massive stars do not rotate at solar metallicity is not a solid one and can be dismissed, being also in contradiction with the observational data.
\\In general, our results confirm that massive stars should rotate faster at low metallicity, but also that their rotational velocities are more scattered going toward lower metallicities. This is a new result which is not present in the literature.
\\It is interesting to compare the behaviour we found for the rotational velocity in massive stars (see Fig.~\ref{fig:5}) with the one assumed in \citet{Prantzos18}, who employed rotating massive stars in a homogeneous Galactic chemical evolution model. In both cases, the yields from \citet{Lim18} are used for RMSs, but here we obtained intermediate velocities by interpolation between the existing grids, while \citet{Prantzos18} changed the fractional contribution of the three rotational velocities considered. The two functions are very similar, but our study predicts an average velocity at low metallicity which is smaller than the one assumed in  \citet{Prantzos18}, and it rapidly goes to zero increasing the metallicity, while the average velocity in \citet{Prantzos18} approaches an asymptotic value of about 50 km/s. To understand this difference, one should keep in mind that our model reproduces the evolution of the Galactic halo, so results at high metallicity cannot be extrapolated.
\\Finally, we show in Fig.~\ref{fig:6} the results of the stochastic model predicting the evolution of strontium and barium, when assuming that rotation in massive stars is described by a Gaussian curve with $\mu$ and $\sigma$ expressed by functions (\ref{eq:1}) and (\ref{eq:2}) respectively. We see that with this model we can reproduce the observations for Sr and Ba at intermediate metallicity, with a high density of points, but also at lower metallicity, where the more dispersed observations are reproduced by low density predictions of the model. This is the most accurate version of the stochastic model we can produce by fine-tuning the parameters which represent the rotation in massive stars.
\\Using these results, we are able now to repeat the test run in Section~\ref{sec:6.1} to compare NSMs and MRD SNe as site of r-process. For this reason, we run again the stocastic model with yields for RMSs from \citet{Lim18} and rotational velocities calibrated with the parameters found above, but with MRD SNe instead of NSMs as source of r-process, as described in Section~\ref{sec:4}. The model-data comparison method for this model returns an index $L=20932$, which is larger than the one obtained with the same model but using NSMs. However, this result should be treated carefully, since it is possible that calibrating the free parameters directly with MRD SNe in the model can lead to new parameters and a lower value for $L$. Therefore, we use this result as an indication that NSMs are expected to better describe the r-process in our model.

\subsection{Predictions for other heavy elements}
\begin{figure*}
\centering
\footnotesize
\includegraphics[trim={0cm 0 0cm 0cm},clip,width=1.\textwidth]{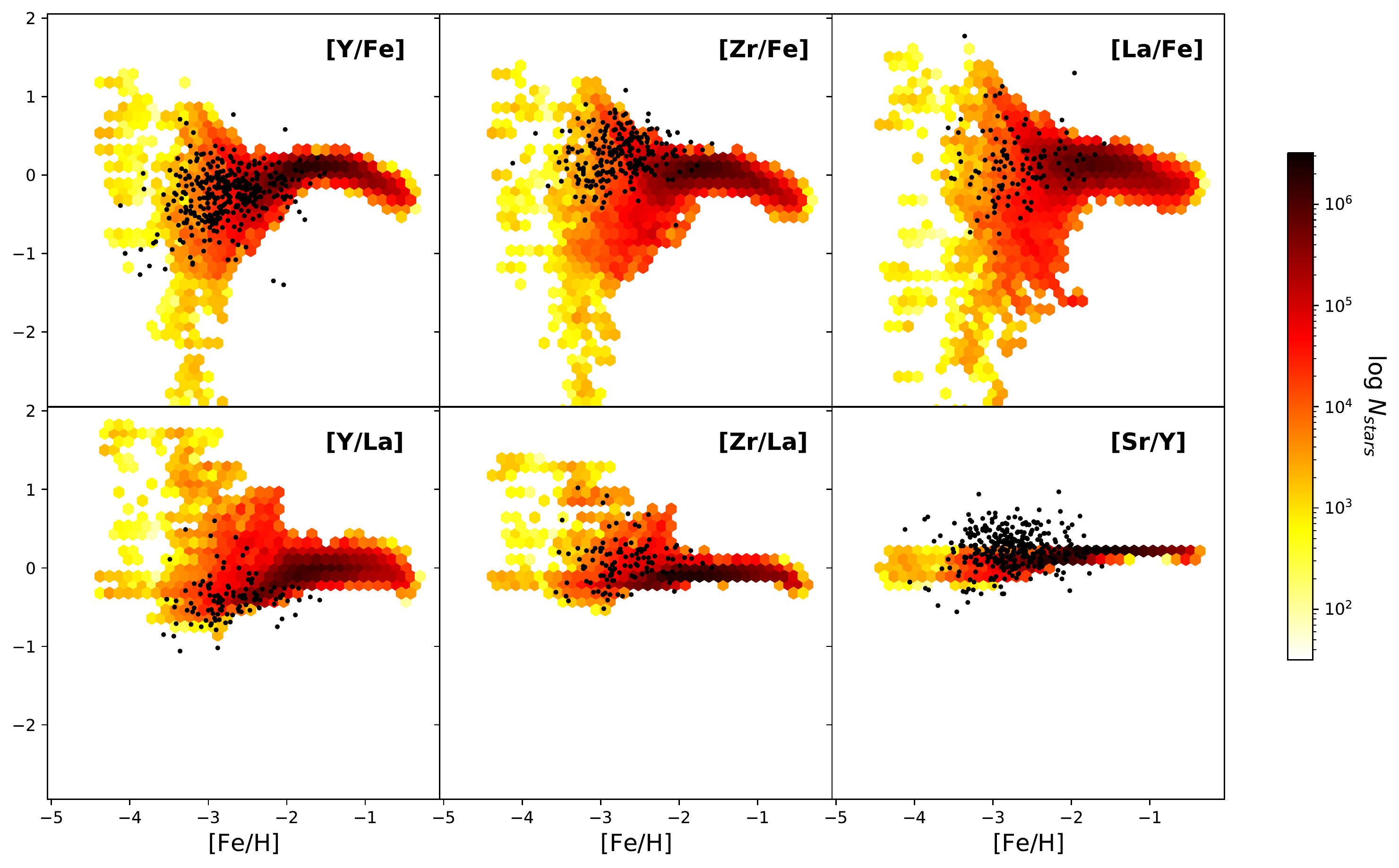}
\caption{Same as Fig.~\ref{fig:6}, with massive star rotational velocities following a Gaussian curve with $\mu$ and $\sigma$ described by functions (\ref{eq:1}) and (\ref{eq:2}) respectively, but for the ratios [Y/Fe], [Zr/Fe], [La/Fe] versus [Fe/H] in the first row, and [Y/La], [Zr/La], [Sr/Y] versus [Fe/H] in the second row.}\label{fig:7}
\end{figure*}

This work is focused on explaining the behaviour of elements strontium and barium through the analysis of rotation in massive stars. However, the stochastic model we employed can process also other neutron capture elements, which we did not use to constrain the free parameters of the model as done with Sr and Ba. It is interesting to see their behaviour in the model we fine-tuned to reproduce at best the observations of strontium and barium.
\\We decide to show here the predictions for elements yttrium, zirconium and lanthanum, since they belong to the first (Y-Zr) and second (La) peaks of s-production, like Sr and Ba. In Fig.~\ref{fig:7} we show different ratios of the three elements versus [Fe/H] for the stochastic model obtained in the previous section, where NSMs were employed as source of r-process and massive stars rotate with velocities following a Gaussian curve with $\mu$ and $\sigma$ described by functions (\ref{eq:1}) and (\ref{eq:2}) respectively, with yields taken from \citet{Lim18}. 
\\As we can see, the stochastic model which is fine-tuned for Sr and Ba is able to correctly reproduce also the observed evolution of [Y/Fe], [Zr/Fe] and [La/Fe], although it predicts some stars at [X/Fe] $<-1$ not supported by observational evidence. We recall the difficulty of measuring heavy elements with very low abundances, so this could be the effect of an observational bias. On the other hand, from the ratios between neutron capture elements we see that the model can reproduce the spread in [Y/La] and [Zr/La], while [Sr/Y] displays a smaller spread since the two elements are produced in similar ratios by r- and s-process. It can be possible that the adopted r-process yields for Y, obtained from the abundance ratios observed in r-process-rich stars \citep{2008ARA&A..46..241S}, may have been overestimated, bringing [Sr/Y] slightly lower than the data.
\\In general, the behaviour of Y, Zr and La shows that, adopting the model fine-tuned for Sr and Ba, also the evolution of other neutron capture elements can be reproduced, so our results are confirmed. Moreover, these elements can be included for further refinements of the model in future studies.

\section{Conclusions}
In this paper, we studied the effects of stellar rotation on the chemical evolution of elements Sr and Ba, using a stochastic model of the Galactic halo presented by \citet{2010A&A...515A.102C} and based on \citet{Cescutti08} and \citet{2008A&A...479L...9C}. We employed different nucleosynthesis prescriptions for s-process in massive stars \citep{Fris16,Lim18} and for r-process production sites (neutron star mergers from \citealt{2014MNRAS.438.2177M}, \citealt{Cescutti15}; magneto-rotationally driven supernovae from \citealt{Cescutti14}). 
\\We summarize our conclusions as follows.
\begin{enumerate}
\item We present a new method to compare model results and observational data. An index of comparison $L$ is defined from the likelihood function, computing the values the model assumes over the data points. As we define it, $L$ is lower for models closer to the observations. In this way, according to the likelihood-ratio test, it is possible to compare different versions of the model and identify the best assumptions. Furthermore, $L$ can be used to estimate the value of free parameters in the model. In this case, we apply the maximum likelihood method, sampling the parameter space and building $L$ as a function of these parameters. From the behaviour of $L$, we obtain the estimates, the errors and the correlations between parameters.
\item Using NSMs or MRD SNe as site of r-process in our stochastic model, with yields for RMSs from \citet{Fris16}, produces very similar results in Sr and Ba abundances. Having applied our method for model-data comparison, we find that employing NSMs produces a model which is closer to the observational data. From this result, we expect that assuming NSMs better describes the r-process in our model.
\item In order to reproduce the observations for Sr and Ba, we studied rotation in massive stars, which contribute via s-process to the enrichment, with yields taken from \citet{Lim18}. Assuming that rotational velocities of massive stars follow a Gaussian probability distribution, we found that the Gaussian $\mu$ and $\sigma$ are dependent on the stellar metallicity and are well described by the exponentially decreasing function (\ref{eq:1}) for $\mu$ and the linear function (\ref{eq:2}) for $\sigma$. The free parameters which characterize these functions were estimated according to the method of model-data comparison previously introduced, assuming that massive stars rotate faster at lower metallicities, and have rotational velocities smaller than 450 $km/s$. With these assumptions, the chemical evolution of Sr and Ba is well reproduced.
\item We analysed the predictions for heavy elements Y, Zr and La in the stochastic model fine-tuned to reproduce the evolution of Sr and Ba. We see that the model is able to reproduce also the evolution of the abundance ratios involving Y, Zr and La. It can be a good idea to include also these elements to constrain the free parameters of the model, which in this way would be able to correctly reproduce the evolution of more neutron capture elements. 
\end{enumerate}
Finally, we would like to underline the importance of having developed and applied a method of model-data comparison for the first time to a stochastic chemical evolution model. This method, employing a completely objective and automatic algorithm, is able to determine which assumptions produce the best results and can estimate the values of free parameters. More in general, this method can have a wide range of applications, in addition to the one presented in this work.

\section*{Acknowledgements}
FM acknowledges funds from University of Trieste (Fondo per la Ricerca d'Ateneo - FRA2016). RH acknowledges support from the World Premier International Research Center Initiative (WPI Initiative), MEXT, Japan. GC and RH acknowledge support from the ChETEC COST Action (CA16117), supported by COST (European Cooperation in Science and Technology). AS acknowledges support from the PRIN-MIUR 2015W7KAWC grant, the INFN INDARK grant, the ERC-StG ‘ClustersXCosmo’ grant agreement 716762, and the FARE-MIUR grant ‘ClustersXEuclid’ R165SBKTMA.
This work has been partially supported by the Italian grants ``Premiale 2015 MITiC'' (P.I. B. Garilli) and ``Premiale 2015 FIGARO'' (P.I. G. Gemme).
\\We acknowledge the computing centre of INAF - Osservatorio Astronomico di Trieste, under the coordination of the CHIPP project, for the availability of computing resources and support \citep{2020arXiv200201283T}.

\section*{Data availability}
The data underlying this article will be shared on reasonable request to the corresponding author.




\bibliographystyle{mnras}
\bibliography{article} 


\bsp	
\label{lastpage}
\end{document}